\documentclass[12pt]{iopart}

\begin{document}

\maketitle

\title{Gauge Functions in Classical Mechanics: From Undriven to 
Driven Dynamical Systems}

\author{Z. E. Musielak, L. C. Vestal, B. D. Tran and T. B. Watson}
\address{Department of Physics, The University of Texas at 
Arlington, Arlington, TX 76019, USA}
\ead{zmusielak@uta.edu}

\begin{abstract}
Novel gauge functions are introduced to non-relativistic classical mechanics 
and used to define forces.  The obtained results show that the gauge functions directly 
affect the energy function and that they allow converting an undriven physical system 
into a driven one.  This is a novel phenomenon in dynamics that resembles the role of 
gauges in quantum field theories.
\end{abstract}


\section{Introduction} 

\label{intro}

The background space and time of non-relativistic Classical Mechanics (CM) 
is described by the Galilean metrics $ds_1^2 = dt^2$ and $ds_2^2 = dx^2 
+ dy^2 + dz^2$, where $t$ is time and $x$, $y$ and $z$ are Cartesian 
coordinates associated with an inertial frame of reference [1].  The metrics 
are invariant with respect to rotations, translations and boots, which form 
the Galilean transformations.  In Newtonian dynamics, the Galilean 
transformations induce a gauge transformation [2], which is called the 
Galilean gauge [3]. The presence of this gauge guarantees that the 
Newton's law of inertia is invariant with respect to the Galilean 
transformations but it also shows that its Lagrangian is not [2,3]. 

A method to remove this gauge was recently proposed [4], and the process 
involves the so-called gauge functions, whose nature and origin are different 
than the Galilean gauge; in other words, the Galilean gauge and the gauge 
functions are different phenomena in CM.  One physical property of these 
functions is that they can be used to remove the unwanted Galilean gauge 
and make the Lagrangian Galilean invariant [4].  The main objective of this 
paper is to demonstrate that these gauge functions can also be used to 
introduce forces into otherwise undriven dynamical systems.   

Different gauge transformations are known in CM and they lead to infinite 
gauge potentials, which in the zero-order become the electromagnetic 
potentials, and in the first-order are identified as the electromagnetic 
and gravitational potentials [5,6].  Gauge transformations in the Lagrangian 
and Hamiltonian formalism of CM, and the resulting diffeomorphism-induced 
gauge symmetries in CM, were also investigated [7], with applications to 
General Relativity.  However, these gauge transformations and their studies 
are not relevant to the gauge functions described in this paper.  
 
In this paper, we generalize the gauge functions derived in [4], and use
them to account for external forces acting on a dynamical system.  We
present a general method to find these gauge functions and apply them 
to simple (linear, undamped, undriven and one-dimensional) oscillators, 
with the purpose to demonstrate how such undriven oscillators can be 
converted into driven ones.   It is suggested that the presented method 
can be applied to other dynamical systems and this gauge function-introduced 
forces may give more physical insight into the connection between forces 
in CM and gauge-introduced interactions in QFT [8].  

For the simple oscillators, the independent variable $t$ is time and the 
dependent variable $x (t)$ is a displacement.  Let $\hat D = {{d^2} / 
{dt^2}} + c$ be an linear differential operator, with $c$ being a constant 
whose value may change from one dynamical system to another, and 
let $\mathcal{Q}$ be a set of all ODEs of the form $\hat D x(t) = 0$; 
depending on the physical meaning of $x(t)$ and $c$, the ODEs of 
$\mathcal{Q}$ may describe different oscillators, including pendulums. 
General solutions of these ODEs are well-known and can be written as 
$x(t) = c_1 x_1(t) + c_2 x_2 (t)$, where $c_1$ and $c_2$ are integration 
constants, and $x_1 (t)$ and $x_2(t)$ are the solutions given in terms 
of the elementary functions [9,10].  

The Lagrangian formalism is established for the ODEs of $\mathcal{Q}$.  
The formalism has always played an important role in obtaining equations 
of motion of dynamical systems [10].  For the conservative dynamical 
systems, the existence of Lagrangians is guaranteed by the Helmholtz 
conditions [11], which can also be used to derive the Lagrangians.  The 
procedure of finding the Lagrangians is called the inverse (or Helmholtz) 
problem of calculus of variations and there are different methods to 
solve this problem [12,13].  We solve the Helmholtz problem and find 
two families of Lagrangians that are classified as primary and general.  
Within each family, two separate classes of Lagrangians are considered, 
namely, standard and null Lagrangians.   

For standard Lagrangians (SLs), the kinetic and potential energy like 
terms and the term with the square of dependent variable are easily 
identified [10,12,13], and these Lagrangians have been known since 
the original work of Lagrange in the 18th Century.   On the other hand, 
null (or trivial) Lagrangians (NLs) contain neither the kinetic nor 
potential energy like terms, and they make the Euler-Lagrange (E-L) 
equation to vanish identically.   Moreover, NLs can also be expressed 
as the total derivative of a scalar function [14,15], which is called a 
gauge function [3].  Our main objective is to obtain the gauge 
functions for the constructed NLs for the ODEs of $\mathcal{Q}$.

The fact that the NLs and their gauge functions can be omitted when 
the original equations are derived is obvious (e.g., [2,3]); however, 
it is also commonly recognized that the NLs are important in studies of 
symmetries of Carath\'eodory's theory of fields of extremals and in integral 
invariants [15,16].  There is a large body of literature on the NLs and on their 
mathematical applictaions (e.g., [17-21]).  Moreover, the NLs play an important 
role in studies of elasticity, where they physically represent the energy density 
function of materials [22,23], and making Lagrangians invariant in the Galilean 
invariant theories [4].    

The main goals of this paper are: (i) construction of the SLs and NLs, and 
the gauge functions corresponding to the NLs; (ii) using these gauge functions 
to determine the energy function and define forces; (iii) deriving new SLs that 
give the equation of motion with the forces; (iv) identifying the gauge functions 
that can be used to define forces in CM; and (v) using the gauge functions to 
convert an undriven oscillator into a driven one.  The presented approach is 
self-consistent and it shows that introducing the gauge functions into CM is 
the equivalent of defining the time-dependent driving forces. 

The outline of the paper is as follows: in Section 2, the Principle of Least 
Action and Lagrangians are described; Section 3 deals with the Lagrangian 
formalism for the considered ODEs and the gauge functions are also derived; 
in Section 4, the energy function for the gauge functions, new definition of 
forces, and the resulting inhomogeneous equations of motion for oscillators 
with different forces are presented and discussed; finally, Section 5 gives our 
conclusions.

\section{Principle of Least Action and Lagrangians}

The Lagrange formalism deals with a functional $\mathcal{A} [x(t)]$, 
where is $A$ is the action and $x(t)$ is an ordinary and smooth function 
to be determined.  Typically $\mathcal{A} [x(t)]$ is given by  an integral 
over a smooth function $L (\dot x, x, t)$ that is called Lagrangian and 
$\dot x$ is a derivative of $x$ with respect of $t$.  The integral defined 
in this way is mathematical representation of the Principle of Least Action 
or Hamilton's Principle [24], which requires that $\delta \mathcal{A} = 0$, 
where $\delta$ is the variation known also as the functional (Fr\'echet) 
derivative of $\mathcal{A} [x(t)]$ with respect to $x(t)$.  Using $\delta 
\mathcal{A} = 0$, the E-L equation is obtained, and this equation is a 
necessary condition for the action to be stationary (to have either a 
minimum or maximum or saddle point).  

We solve the inverse problem of the calculus of variations for the ODEs 
of $\mathcal{Q}$ and find their SLs and NLs; the validity of the Helmholtz 
conditions [6] for these Lagrangians is also discussed.  Different methods 
were previously developed to determine the SLs for different ODEs [25-34] 
and some of these methods [25,26] will be used in the next section.  Based
on the original work of Lagrange in the 18th Century, the SLs contain the 
difference between the kinetic and potential energy like terms, which in 
here will be represented by {\it the difference between the square of the 
first order derivative of the dependent variable and the term with the 
square of dependent variable} [10,12,13].   

On the other hand, the NLs contain neither kinetic nor potential energy 
like terms but instead {\it they depend on terms with mixed dependent 
variable and its derivative} [14,20], {\it and terms with mixed dependent 
variable (or its derivative) with the independent variable, and also 
terms that depend only on the dependent variable}.  The derived NLs
are new and they are restricted to the lowest order in the dependent 
variable.  For any NL, the E-L equation identically vanishes, and any 
NL can be expressed as the total derivative of their gauge functions.
Our main results are novel gauge functions obtained for the NLs and 
their role in converting undriven dynamical systems into driven ones.

\section{Lagrangians and gauge functions}

\subsection{Standard and null Lagrangians}

Using the definition of SLs given in Section 2, they can be written in 
the following form 
\begin{equation}
L_{s} [\dot x(t), x(t)] = {1  \over 2} \left [ \alpha \left ( \dot x (t) \right )^2
+ \beta x^2 (t) \right ]\ ,
\label{S3eq1}
\end{equation}
where the coefficients $\alpha$ and $\beta$ are either constants or 
functions of time; most SLs obtained here are already known [10,12,13].
The NLs are defined in Section 2, and let us point out that the derived 
NLs are new.  We now follow [4] to show how NLs with constant 
coefficients can be constructed and then generalize this approach 
in Section 3.3. 

Let $L_{m} [\dot x(t), x(t)]$ be a mixed Lagrangian of the dependent 
and independent variables given by 
\begin{equation}
L_{m} [\dot x(t), x(t), t] = C_1 \dot x (t) x (t) + C_2 \dot x (t) t + 
C_3 x (t) t\ ,
\label{S3eq2}
\end{equation}
and $L_{f} [\dot x(t), x(t)]$ be a Lagrangian of the single dependent 
variable written as 
\begin{equation}
L_{f} [\dot x(t), x(t)] = C_4 \dot x (t) + C_5 x (t) + C_6\ ,
\label{S3eq3}
\end{equation}
where $C_1$, $C_2$, $C_3$, $C_4$, $C_5$ and $C_6$ are arbitrary 
constants.  However, with $x (t)$ being a displacement of harmonic 
scillators and $t$ being time, the constants must have different physical 
dimensions to get the same dimensions of $L_{m} [\dot x(t), x(t),t]$ 
and $L_{f} [\dot x(t), x(t)]$ as that of $L_{s} [\dot x(t), x(t)]$.

We define $\hat {EL}$ to be the E-L equation operator and take 
$\hat {EL} (L_m + L_f) = 0$, which is required for $L_{n} [\dot x(t), 
x(t), t] = L_{m} [\dot x(t), x(t), t] + L_{f} [\dot x(t), x(t)]$ to become the 
null Lagrangian.  This is true if, and only if, $C_3 = 0$ and $C_5 = C_2$.  
Then, the null Lagrangian can be written [4] as 
\begin{equation}
L_{n} [\dot x(t), x(t), t] = \sum_{i = 1}^{4} L_{ni} [\dot x(t), x(t), 
t]\ ,
\label{S3eq4}
\end{equation}
where $i$ = 1, 2, 3 and 4, and the partial NLs are given by $L_{n1} [\dot x(t), 
x(t)] = C_1 \dot x (t) x (t)$, $L_{n2} [\dot x(t), x(t), t] = C_2 [ \dot x (t) t + 
x (t) ]$, $L_{n3} [\dot x(t)] = C_4 \dot x (t)$ and $L_{n4} = C_6$; with 
$L_{n2} [\dot x(t), x(t), t]$ being the only partial null Lagrangian that depends
explicitly on $t$.   Note that these partial null Lagrangians are constructed to 
lowest orders of the dynamic variable $x (t)$.

Since $L_{n} [\dot x(t), x(t), t] = d \Phi_p / d t$, we may write the gauge function 
$\Phi_p (t)$ [4] as  
\begin{equation}
\Phi_p (t) = \sum_{i = 1}^4 \phi_{pi} (t)\ , 
\label{S3eq5}
\end{equation}
where the partial gauge functions $\phi_{pi} (t)$ correspond the partial null Lagrangians 
$L_{ni} [\dot x(t), x(t)]$, and they are defined as $\phi_{p1} (t) = C_1 x^2 (t) / 2$, 
$\phi_{p2} (t) = C_2 x (t) t$, $\phi_{p3} (t) = C_4 x (t)$ and $\phi_{p4} (t) = C_6 t$.  

We now use the above results to derive the SLs, NLs and gauge functions for 
the ODEs of $\mathcal{Q}$.

\subsection{Primary Lagrangians and gauge functions}

We consider the ODEs of $\mathcal{Q}$ and write them in their explicit form
\begin{equation}
\ddot x (t) + c x (t) = 0\ ,
\label{S3eq6}
\end{equation}
where $c$ may be any real number.  Let us define the following primary 
Lagrangian  
\begin{equation}
L_{p} [\dot x(t), x(t), t] = L_{ps} [\dot x(t), x(t)] + L_{pn} [\dot x(t), 
x(t), t]\ ,
\label{S3eq7}
\end{equation}
where the primary standard Lagrangian (with $\alpha = 1$ and $\beta = 
- c$ in Eq. \ref{S3eq1}) is given by 
\begin{equation}
L_{ps} [\dot x(t), x(t)] = {1  \over 2} \left [ \left ( \dot x (t) \right )^2 
- c x^2 (t) \right ]\ ,
\label{S3eq8}
\end{equation}
and the primary null Lagrangian $L_{pn} [\dot x(t), x(t)]$ is equal to
$L_{n} [\dot x(t), x(t)]$ (see Eq. \ref{S3eq4}) with the same partial 
NLs.  In addition, the primary gauge function $\Phi_p (t)$ is given by 
Eq. (\ref{S3eq5}) with the same partial gauge functions.

\subsection{General Lagrangians and gauge functions}

The above results can be generalized by writing the Lagrangian 
given by Eq. (\ref{S3eq1}) in the following form
\begin{equation}
L_{s} [\dot x(t), x(t)] = {1  \over 2} \left [ \alpha (t) \left ( \dot x (t) \right )^2
+ \beta (t) x^2 (t) \right ]\ ,
\label{S3eq9}
\end{equation}
where $\alpha (t)$ and $\beta (t)$ are continuous and differentiable 
functions.  Substituting this Lagrangian to the E-L equation, we find 
$\alpha (t) = C_o$ and $\beta (t) = - C_o c$, where $C_o$ is an 
intergration constant.  Then, the general standard Lagrangian can 
be written as 
\begin{equation}
L_{gs} [\dot x(t), x(t)] = {1  \over 2} C_o \left [ \left ( \dot x (t) \right )^2
- c x^2 (t)] \right ]\ .
\label{S3eq10}
\end{equation}
This Lagrangian can be reduced to the primary standard Lagrangian 
if $C_o = 1$ and it can also be used to define the following general 
Lagrangian 
\begin{equation}
L_{g} [\dot x (t), x(t), t] = L_{gs} [\dot x (t), x(t), t] + 
L_{gn} [\dot x (t), x(t), t]\ ,
\label{S3eq11}
\end{equation}
where the general null Lagrangian is
\begin{equation}
L_{gn} [\dot x (t), x(t), t] = \sum_{i = 1}^4 L_{gni} [\dot x (t), x(t), t]\ ,
\label{S3eq12}
\end{equation}
with $\hat {EL} (L_{gn}) = 0$ and $L_{gni} [\dot x (t), x(t), t]$
being its partial components.  To determine the partial null Lagrangians, 
we generalize the primary gauge functions $\phi_{pi} (t)$ given 
below Eq. (\ref{S3eq5}) by replacing their constant coefficients by 
functions of the independent variable $t$.  Denoting the general 
gauge functions as $\phi_{gi} (t)$, we obtain
\begin{equation}
\phi_{g1} (t) = {1 \over 2} f_1 (t) x^2 (t)\ ,
\label{S3eq13}
\end{equation}
\begin{equation}
\phi_{g2} (t) = f_2 (t) x (t) t\ ,
\label{S3eq14}
\end{equation}
\begin{equation}
\phi_{g3} (t) = f_4 (t) x (t)\ ,
\label{S3eq15}
\end{equation}
and
\begin{equation}
\phi_{g4} (t) = f_6 (t) t\ ,
\label{S3eq16}
\end{equation}
where $f_1 (t)$, $f_2 (t)$, $f_4 (t)$ and $f_6 (t)$ are continuous 
and differentiable functions to be determined. 

Then, we take the total derivative of these partial gauge functions
and obtain the following partial Lagrangians
\begin{equation}
L_{gn1} [\dot x(t), x(t), t] = \left [ f_1 (t) \dot x (t) + {1 \over 2} 
\dot f_1 (t) x (t) \right ] x (t)\ ,
\label{S3eq17}
\end{equation}
\begin{equation}
L_{gn2} [\dot x(t), x(t), t] = \left [ \left ( f_2 (t) \dot x (t) + 
\dot f_2 (t) x (t) \right ) t + f_2 (t) x \right ]\ ,
\label{S3eq18}
\end{equation}
\begin{equation}
L_{gn3} [\dot x(t), x(t), t] = \left [ f_4 (t) \dot x (t)  + \dot f_4 (t) 
x (t) \right ]\ ,
\label{S3eq19}
\end{equation}
and
\begin{equation}
L_{gn4} [\dot x(t), x(t), t] = \left [ \dot f_6 (t) t + f_6 (t) \right ]\ ,
\label{S3eq20}
\end{equation}
which can be added together to obtain the general null Lagrangian
(see Eq. \ref{S3eq12}).  This Lagrangian depends on four functions 
that must be continous and differentiable but otherwise arbitrary.  
Specification of initial conditions for physical problems would set up 
constraints on these functions, however, in this paper the functions 
are kept arbitrary for reasons explained in Sect. 4.  

The general null Lagrangian reduces to the primary null Lagrangian 
when $f_1 (t) = C_1$, $f_2 (t) = C_2$, $f_4 (t) = C_4$ and $f_6 
(t) = C_6$.

\subsection{Discussion of Lagrangians and gauge functions}

The obtained results show that the SLs and NLS can be found for
the ODEs of $\mathcal{Q}$ by solving the inverse problem of the 
calculus of variations.  The existence of these Lagrangians must be 
validated by the Helmholtz conditions [11].  There are three original 
Helmholtz conditions and it is easy to verify that all Lagrangians 
constructed for $\hat D x (t) = 0$ obey these conditions, which 
means that the SLs do exist for undamped (conservative) systems 
[5-8].  Let us also point out that the existence of NLs is not affected 
by the Helmholtz conditions because these Lagrangians have no 
effects on the derivation of the original equations.

We derived the primary and general SLs and NLs for the ODEs of 
$\mathcal{Q}$.  Most obtained SLs are already known and they 
are generated as a byproduct of our procedure of deriving the 
NLs, which are new for the considered equations.  For each 
null Lagrangian, we found its corresponding gauge function.  
The general Lagrangians depend on four functions that must be 
continuous and differentiable, and must satisfy initial conditions 
of a specific physical problem.  If the functions are assumed to 
be constants, the primary NLs are obtained.  Since the 
functions are arbitrary, many different NLs can be obtained 
by choosing different forms of these functions.   

It was previously demonstrated that a Lagrangian is null if, and 
only if, it can be represented as the total derivative of a scalar 
function of the system variables [14].  If this function exists, the 
resulting transformation is called the gauge transformation and 
the function is known as a gauge function [3,4].  The results 
presented here demonstrate that this definition is valid for the 
ODEs of $\mathcal{Q}$, and that for all these equations the 
gauge functions exist.  The presented gauge functions were 
derived here only for the ODEs with the constant coefficients; 
notably, the derivations can also be extended to the ODEs with 
non-constant coefficients and first attempts in finding such gauge 
functions are described in [33,34]. 

Since the obtained NLs are given as total derivatives of scalar functions,
they can be omitted from the Lagrangians when the original equations 
are derived from the E-L equation [2,3,35].  However, the purpose of 
this paper is to determine the NLs and derive the corresponding gauge 
functions, which are then used to convert undriven for oscillators (or
pendulums) into driven systems.  In other words, the gauge functions 
are used to introduce forces to CM, which is a new phenomenon.

\section{Application: from undriven to driven oscillators}

\subsection{Primary gauge and energy functions}

Let us consider a harmonic {\it oscillator} and identify $x (t)$ with 
its displacement variable.  The equation of motion of the oscillator 
is $\hat D x (t) = 0$ with $c = k / m$, where $k$ is a spring 
constant and $m$ is mass.  The characteristic frequency of the 
oscillator is then $\omega_o = \sqrt{c} = \sqrt{k / m}$, and 
the equation of motion can be written as  
\begin{equation}
\ddot x (t) + \omega_o^2 x (t) = 0\ .
\label{S4eq1}
\end{equation}
It must be noted that Eq. (\ref{S4eq1}) also describes a linear 
and undamped {\it pendulum} if $x (t)$ is replaced by $\theta 
(t)$, where $\theta (t)$ is an angle of the pendulum, and 
$\omega_o$ is replaced by the pendulum characteristic 
frequency $\omega_p = \sqrt{c} = \sqrt{g / L}$, where $g$ 
is gravitational acceleration and $L$ is length of the pendulum.
With these replacements, the results presented below for the 
oscillator are also valid for the pendulum. 

According to Eq. (\ref{S3eq7}), the primary Lagrangian $L_{p} 
[\dot x(t), x(t)]$ for these harmonic oscillators can be written as 
\begin{equation}
L_{p} [\dot x(t), x(t)] = L_{ps} [\dot x(t), x(t)] + {{d \phi_p} 
\over {dt}} 
\ ,
\label{S4eq2}
\end{equation}
where the primary standard Lagrangian is given by 
\begin{equation}
L_{ps} [\dot x(t), x(t)] = {1  \over 2} \left [ \left ( \dot x (t) 
\right )^2 - \omega_o^2 x^2 (t) \right ]\ ,
\label{S4eq3}
\end{equation}
and the primary gauge function $\Phi_{p}$ is
\begin{equation}
\Phi_p (t) = \sum_{i = 1}^4 \phi_{p i} (t)\ , 
\label{S4eq4}
\end{equation}
and the partial primary gauge functions are: 
\begin{equation}
\phi_{p1} = {1  \over 2} C_1 x^2 (t)\ ,
\label{S4eq5}
\end{equation}
\begin{equation}
\phi_{p2} = C_2 x (t) t\ ,
\label{S4eq6}
\end{equation}
\begin{equation}
\phi_{p3} = C_4 x (t)\ ,
\label{S4eq7}
\end{equation}
and 
\begin{equation}
\phi_{p4} = C_6 t\ .
\label{S4eq8}
\end{equation}
Note that the total derivative of each one of these partial
gauge functions gives no contribution to the resulting 
equation of motion.  However, these gauge functions
may be used to impose Galilean invariance of SLs [4].

Since the gauge functions $\phi_{p2}$ and $\phi_{p4}$ 
depend explicitly on time $t$, the resulting primary null 
Lagrangian is also a function of time.  This requires that 
the primary energy function, $E_{p}$, is calculated 
[36,37] using 
\begin{equation}
E_{p} [\dot x(t), x(t)] = \dot x {{\partial L_{p}} \over {\partial 
\dot x}} - L_{p} [\dot x(t), x(t)]\ ,
\label{S4eq9}
\end{equation}
which gives
\begin{equation}
E_{p} [\dot x(t), x(t)] = {1  \over 2} \left [ \left ( \dot x (t) \right )^2 
+ \omega_o^2 x^2 (t) \right ] - \left [ C_2 x + C_6 \right ]\ ,
\label{S4eq10}
\end{equation}
with the first two terms on the RHS representing the energy function 
$E_{ps}$ for the primary standard Lagrangian and the other two terms 
corresponding to the primary energy function $E_{pf}$ for the primary 
gauge function, so that $E_{p} = E_{ps} + E_{pf}$.  

In general, $E_{p} \neq E_{tot}$, with  $E_{tot} = E_{ps} = H_{ps}$, 
where $E_{tot}$ is the total energy of system and $H_{ps}$ is its 
Hamiltonian, corresponding to the primary standard Lagrangian, and 
given by $H_{ps} = E_{p} - E_{pf}$ or  
\begin{equation}
H_{ps} [\dot x(t), x(t)] =  {1  \over 2} \left [ \dot x^2 (t) + \omega_o^2 
x^2 (t) \right ]\ .
\label{S4eq11}
\end{equation}
Using the Hamilton equations, the equation of motion for the harmonic 
oscillator given by Eq. (\ref{S4eq1}) is obtained.  Similar result is derived
when the total derivative of $E_{p}$ is equal to the negative partial time 
derivative of $L_{p}$ that can be written [36] as 
\begin{equation}
{{d E_{p}} \over {dt}} = - {{\partial L_{p}} \over {\partial t}}\ ,
\label{S4eq12}
\end{equation}
which again gives Eq. (\ref{S4eq1}).  It must be noted that $E_{p}$ is 
a conserved quantity and that $E_{p} \neq E_{tot}$.   This shows that 
the equation of motion of the harmonic oscillator is also obtained when 
the energy function is used instead of the primary Lagrangian $L_{p}$ 
or the Hamiltonian $H_{ps}$.  

The above results show that among the four primary gauge functions, 
$\phi_{p1}$, $\phi_{p2}$, $\phi_{p3}$ and $\phi_{p4}$, the first and 
third do not contribute to the primary energy function, but the second 
and fourth do contribute although each one differently.  The partial 
gauge function $\phi_{p2}$ breaks into two parts and only the part 
that depends on $C_2 x$ contributes to the energy function.  However, 
the partial gauge function $\phi_{p4}$ fully contributes to the energy 
function.  Let us call $\phi_{p2}$ the {\it primary F-gauge function}, 
and $\phi_{p4}$ the {\it primary E-gauge function}.  

The reasons for these names follows.  First, the term $C_2 x$ 
represents energy if, and only if, the coefficient $C_2$ is a constant 
acceleration, or a constant force per mass, so that $C_2 x$ is work 
done by this force on the system.  This clearly shows that the primary 
partial gauge function $\phi_{p2}$ can be used to introduce forces 
that cause the constant acceleration.  Second, the primary partial 
gauge function $\phi_{p4}$ introduces a constant energy shift in the system.  

Let us define $F_c = C_2$, where $F_c$ represents a constant acceleration 
or constant force per mass.   Similarly, $E_c = C_6$ is a constant energy 
shift that could be caused by the force.  Then, the primary energy function 
can be written 
as
\begin{equation}
E_{p} [\dot x(t), x(t)] = {1  \over 2} \left [ \left ( \dot x (t) \right )^2 
+ \omega_o^2 x^2 (t) \right ] - \left [ F_c x + E_c \right ]\ .
\label{S4eq13}
\end{equation}
This demonstrates that some gauge functions can be used to introduce 
external forces that drive the system but other gauge functions may 
either generate a shift of the total energy of the system, or simply have 
no effect on the system.  In other words, only {\it gauge functions that 
depend explicitly on time} may be used to introduce forces in CM.  
These are new phenomena caused exclusively by including the gauge 
functions into CM.

\subsection{General gauge and energy functions}
 
The above results can be now extended to the general standard and 
null Lagrangians and their gauge functions with application to a 
harmonic oscillator and pendulum.  According to Eqs (\ref{S3eq10} 
through \ref{S3eq12}), the general Lagrangian for the oscillator 
can be written as 
\begin{equation}
L_{g} [\dot x (t), x(t), t] = L_{gs} [\dot x (t), x(t), t] + 
L_{gn} [\dot x (t), x(t), t]\ ,
\label{S4eq14}
\end{equation}
where the general standard and null Lagrangian are 
\begin{equation}
L_{gs} [\dot x(t), x(t)] = {1  \over 2} C_o \left [ \left ( \dot x (t) 
\right )^2 - \omega_o^2 x^2 (t)] \right ]\ ,
\label{S4eq15}
\end{equation}
and
\begin{equation}
L_{gn} [\dot x (t), x(t), t] = \sum_{i = 1}^4 {{d \phi_{gi}} \over 
{dt}}\ ,
\label{S4eq16}
\end{equation}
with the partial gauge functions $\phi_{gi} (t)$ being given by Eqs 
(\ref{S3eq13}) through (\ref{S3eq16}). 

The general energy function, $E_{g} [\dot x(t), x(t)]$, can be calculated 
by substituting $L_{gn} [\dot x (t), x(t), t]$ into Eq. (\ref{S4eq9}), which 
gives  
\begin{equation}
E_{g} [\dot x(t), x(t)] = E_{gs} [\dot x(t), x(t)] + E_{gf} [\dot x(t), x(t)]\ , 
\label{S4eq17}
\end{equation}
where the general energy function for the general standard Lagrangian is 
\begin{equation}
E_{gs} [\dot x(t), x(t)] = {1  \over 2} C_o \left [ \left ( \dot x (t) \right )^2 
+  \omega_o^2 x^2 (t) \right ]  \ . 
\label{S4eq18}
\end{equation}
and the general energy function for the general gauge function can be 
written as 
\[
E_{gf} [\dot x(t), x(t)] = - \left [ {1 \over 2} \dot f_1 (t) x^2 (t) + 
\dot f_2 (t) x (t) t \right ]   
\]
\begin{equation}
\hskip0.5in - \left [ \left ( f_2 (t) + \dot f_4 (t) \right ) x (t) + 
f_6 (t) + \dot f_6 (t) t \right] \ . 
\label{S4eq19}
\end{equation}

Since $E_{gs} = H_{gs} = E_{tot}$, then $H_{gs} = E_{g} - E_{gf}$ 
and, as expected, when $H_{gs} $ is substituted into the Hamilton 
equations, the equation of motion for the harmonic oscillator (see Eq. 
\ref{S4eq1}) is obtained.  The same equation of motion is derived when 
the total derivative of $E_{g}$ is equal to the negative partial time 
derivative of $L_{g}$ (see Eq. \ref{S4eq12}).

The obtained results show that the general gauge functions, $\phi_{g1}$ 
and $\phi_{g3}$ also contribute to the general energy function, in addition, 
to the $\phi_{g2}$ and $\phi_{g4}$ contributions.  We generalize the previous 
definitions and now call $\phi_{g2}$ the {\it general F-gauge function}, and 
$\phi_{g4}$ the {\it general E-gauge function}.  However, no special names 
are given to the gauge functions $\phi_{g1}$ and $\phi_{g3}$, and only 
their contributions to forces is shown below.

We may define the following functions: $F (t, x) = [ f_2 (t) + \dot f_2 (t) t + 
\dot f_4 (t)] x(t)$ and $G (t) = f_6 (t) + \dot f_6 (t) t$ and see that all 
gauge functions contribute to them.  Using these definitions, we write
\[
E_{g} [\dot x(t), x(t)] = {1  \over 2} \left [ \left ( \dot x (t) \right )^2 
+ \omega_o^2 \left ( 1 - {{\dot f_1 (t)} \over {\omega_o^2}} \right ) 
x^2 (t) \right ] 
\]
\begin{equation}
\hskip0.75in - \left [ F (t, x) + G (t) \right ]\ ,
\label{S4eq20}
\end{equation}
which shows that the gauge functions allow us to introduce two functions, 
one that depends linearly on displacement but is arbitrary in time, and the 
other that is an arbitrary function of time only.  This general formula for 
the energy function may be further simplified by taking $f_1 (t) = C_1$ 
= const, which means that the shift of the potential energy is not time-dependent 
and remains constant all the time.  Then, the general energy function becomes   
\begin{equation}
E_{g} [\dot x(t), x(t)] = {1  \over 2} \left [ \left ( \dot x (t) \right )^2 
+ \omega_o^2 x^2 (t) \right ] - \left [ F (t, x) + G (t) \right ]\ ,
\label{S4eq21}
\end{equation}
and the function $F (t, x)$ reduces to the primary energy function
if $F (t, x) = F_c x (t)$, and the function $G (t)$ becomes $E_c$ 
(see Eq. \ref{S4eq13}).

\subsection{Time-dependent forces}

Having obtained the general energy function $E_{g} [\dot x(t), x(t)]$,
for the equations of motion of undriven oscillators, we now demonstrate 
that these systems can be converted into driven ones.  This can be done 
by adding the extra terms $F (t, x)$ and $G (t)$ to the general standard
Lagrangian.  Let us separate the dependent and independent variables 
in $F (t, x)$ and write $F (t, x) = {\cal {F}} (t) x (t)$.  The result is 
\begin{equation}
L_{g} [\dot x(t), x(t)] =  L_{gs} [\dot x(t), x(t)] + \left [ {\cal {F}} (t) x 
+ G (t) \right ]\ ,
\label{S4eq22}
\end{equation}
where $L_{ps} [\dot x(t), x(t)]$ and $L_{gs} [\dot x(t), x(t)]$ are 
given by Eqs (\ref{S4eq3}) and (\ref{S4eq15}), respectively. 
Substituting $L_{g} [\dot x(t), x(t)]$ into the E-L equation, we
obtain 
\begin{equation}
\ddot x (t) + \omega_o^2 x (t) = {\cal {F}} (t)\ .
\label{S4eq23}
\end{equation}
This equation describes a driven oscillator with ${\cal {F}} (t)$ 
being a time-dependent force. The equation also represents a linear 
undamped pendulum if $x (t)$ is replaced by $\theta (t)$ and $\omega_o$ 
is replaced by $\omega_p$.  In a special case of the primary null 
Lagrangian with constant coefficients (see Eq. \ref{S3eq7}), the 
force ${\cal {F}} (t)$ is the constant force $F_c$.  

Let us point out that the resulting inhomogeneous equation of motion is 
also obtained from the Hamilton equations when the energy function is 
used instead of the Hamiltonians $H_{ps} [\dot x(t), x(t)]$ and $H_{gs} 
[\dot x(t), x(t)]$. This is expected as the Hamiltonians represent the total 
energy of the system, which is not conserved but the energy function is 
a constant of motion for the considered driven harmonic oscillator.  This 
shows that our approach is self-consistent and based on the principles of 
CM.  However, by accounting for the gauge functions and by showing their 
relationships to forces, this paper describes a new phenomenon in CM, 
which formally allows converting undriven dynamical systems into 
driven ones.  The converting process can be used for any linear 
dynamical system for whose equation of motion is known.

\section{Conclusions} 

The Lagrangian formalism was established for equations describing different 
undriven dynamical systems by constructing the standard and null Lagrangians, 
and the gauge functions corresponding to the latter.  The gauge functions 
were used to determine the energy function and define forces.  Using these 
forces, new standard Lagrangians were obtained and the equations of motion 
resulting from these Lagrangians were derived.  It was shown that the 
equations of motion are inhomogeneous because of the presence of the 
time-dependent driving forces introduced by the gauge functions, and 
that the same equations can be obtained by using either the energy 
function or the Hamilton equations.  Moreover, the obtained results 
demonstrate that the approach does not allow defining dissipative forces 
that depend on velocities.  It was also pointed out that only some gauge 
functions give the driving forces and those gauge functions were identified 
and discussed.  

The presented approach is self-consistent and it shows that introducing 
the gauge functions into Classical Mechanics is equivalent of finding the
time-dependent driving forces; it must be noted that the gauge functions 
derived in this paper are different the gauges considered before.  The 
obtained results demonstrate that not all gauge functions give forces, 
instead there is only one primary and only one general gauge function 
that introduces the driving forces to Classical Mechanics.  This new 
phenomenon of defining the driving forces in Classical Mechanics by 
the gauge functions, and converting an undriven system into a driven 
one, can be easily generalized to other linear dynamical systems, 
either conservative or non-conservative.  The phenomenon resembles 
the role of gauges in quantum field theories but there are differences 
in the underlying physics that will be investigated separately.

\bigskip\noindent
{\bf Acknowledgments}
We are indebted to two anonymous referees for their valuable suggestions
that have allowed us to improve significantly our original manuscript.  This 
work was supported by  the Alexander von Humboldt Foundation (Z.E.M.). 

\bigskip\noindent
{\bf References}

\nocite{*}

\end{document}